\documentclass[%
reprint,
superscriptaddress,
 amsmath,amssymb,
 aps,
]{revtex4-2}
\usepackage{color}
\usepackage{graphicx}% Include figure files
\usepackage{dcolumn}% Align table columns on decimal point
\usepackage{bm}% bold math
\begin{document}

\preprint{APS/123-QED}

\title{Percolation analysis of spatiotemporal distribution of population in Seoul and Helsinki}% Force line breaks with \\

\author{Yunwoo Nam}
\affiliation{Department of Physics, University of Seoul, Seoul 02504, Republic of Korea}

\author{Young-Ho Eom}%
 \email{yheom@uos.ac.kr}
\affiliation{Department of Physics, University of Seoul, Seoul 02504, Republic of Korea}
\affiliation{Natural Science Research Institute, University of Seoul, Seoul 02504, Republic of Korea}
\affiliation{Urban Big data and AI Institute, University of Seoul, Seoul 02504, Republic of Korea}

%              %  but any date may be explicitly specified

\begin{abstract}
Spatiotemporal distribution of urban population is crucial to understand the structure and dynamics of cities. Most studies, however, have focused on the microscopic structure of cities such as their few most crowded areas. In this work, we investigate the macroscopic structure of cities such as their clusters of highly populated areas. To do this, we analyze the spatial distribution of urban population and its intraday dynamics in Seoul and Helsinki with a percolation framework. We observe that the growth patterns of the largest clusters in the real and randomly shuffled population data are significantly different, and highly populated areas during the daytime are denser and form larger clusters than highly populated areas during the nighttime. An analysis of the cluster-size distributions at percolation criticality shows that their power-law exponents during the daytime are lower than those during the nighttime, indicating that the spatial distributions of urban population during daytime and nighttime fall into different universality classes. Finally measuring the area-perimeter fractal dimension of the collection of clusters demonstrates that the fractal dimensions during the daytime are higher than those during the nighttime, indicating that the perimeters of clusters during the daytime are rougher than those during the nighttime. Our findings suggest that even the same city can have qualitatively different spatial distributions of population over time, and propose a way to quantitatively compare the macrostructure of cities based on population distribution data.

\end{abstract}

% %\keywords{Suggested keywords}%Use showkeys class option if keyword
%                               %display desired
\maketitle

% %\tableofcontents 

\section{\label{sec:level1} Introduction }
Characterizing the spatio-temporal distribution of urban population is important because it not only helps us to understand the structure and dynamics of cities but also provides essential information on the demand for traffic, housing, and infrastructure in cities. Spatial distributions of population in many cities are not uniform but quite heterogeneous~\cite{volpati2018spatial,schwarz2010urban}. A key concept in urban structure, thus, is urban centers or hotspots, usually defined as the most crowded places in cities~\cite{barthelemy2016structure, bettencourt2021introduction}. Urban structure can be classified as either monocentric or polycentric based on the number of centers and as either compact or dispersed based on how closely clustered centers are \cite{hong2022relationship, xu2023urban}. 

Until recently, the registered population based on censuses has been mainly used to understand the spatial distribution of urban population~\cite{volpati2018spatial, sang2024spatial, crosato2021polycentric}. These data are highly reliable as collected by the governments, but they are updated every five to ten years and are based on a residence-based population, which limits its ability to describe the dynamics of urban structure. However, real-time or near-real-time population data collected from mobile phone users have recently become available and can be used to better understand urban structure and dynamics~\cite{dong2020understanding, louail2015uncovering, lee2021spatiotemporal, wang2009understanding, lenormand2014cross, wu2021spatial}. A remarkable work proposed a parameter-free method to identify hotspots in 31 Spanish cities, showed that the number of hotspots in these cities scales sublinearly with their population, and classified these hotspots according to their lifetime~\cite{louail2014mobile}. Mobile phone data also allows one to investigate the urban structure encoded in mobility flows of people in cities with the notion of urban centers~\cite{roth2011structure, bassolas2019hierarchical}. These center-based investigations have focused on the microscopic structure of cities although they have revealed important structural properties of cities. A comprehensive understanding of the structure and dynamics of a city requires an investigation of the macrostructure of the spatial distribution of the city's population such as clusters of highly populated areas.

Percolation can provide an effective way to investigate the macroscopic structure underlying a given spatial distribution of variables of interest from temperature and altitude to congestion and population~\cite{saberi2015recent, stauffer2018introduction, ebrahimabadi2023geometry, zeng2019switch, kwon2023global, cao2020quantifying, ambuhl2023understanding, ali2013percolation, fan2019percolation, sun2021percolation, rozenfeld2011area}. For example, a work that analyzed spatial distribution of temperature in cities with a temperature-based percolation (i.e., occupying sites in order of the highest temperature to the lowest temperature) revealed that urban heat islands are fractal despite the diversity in urban form, composition, and environment across the world~\cite{shreevastava2019emergent}. Another work, using a percolation model to analyze spatial distributions of congestion in two cities, showed that these distributions belong to different universality classes of percolation depending on the commuting patterns of the cities~\cite{ebrahimabadi2023geometry}. Although many previous works on urban structure have focused on center-based approaches, these have limitations in revealing urban macrostructure\cite{dong2020understanding, deng2019detecting, sadewo2021using, li2018did, krehl2018urban}. With a percolation-based framework, we can not only understand urban macrostructure in terms of clusters but can also quantitatively compare macroscopic characteristics of urban structure, for example, using critical exponents and fractal dimensions.

In this work, we investigate the intraday dynamics of the macroscopic structure underlying the spatial distribution of populations in Seoul and Helsinki with a percolation framework. We examine the characteristics of clusters of highly populated areas and the connectivity between these clusters as they are created by occupying areas from high to low population in this framework. We observe that the growth patterns of the largest clusters in the real-world and randomly shuffled population data are significantly different, and highly populated areas during the daytime are denser and form larger clusters than highly populated areas during the nighttime by analyzing the size of the largest clusters, the radius of gyration of occupied areas, and the correlation length of clusters in the data. We investigate the cluster-size distributions at the percolation critical point and observe that their power-law exponents during daytime are lower than during nighttime, indicating that the spatial distributions of urban population during daytime and nighttime belong to different universality classes. We measure the area-perimeter fractal dimension of the collection of clusters and observe that the fractal dimensions during the daytime are higher than those during the nighttime, indicating that the perimeters of high-population clusters during daytime are rougher than during nighttime.

\section{Data And Method}
\subsection{\label{app:subsec}Data}
We use the Seoul Living Population Data (SLPD)~\cite{SLPD} provided by the Seoul Metropolitan Government to analyze the spatio-temporal distribution of population in Seoul. The SLPD is based on mobile phone signal data and tells us the total number of people of a specific area in Seoul at a specific time (an hourly interval). There are 19,153 area units, called ‘jibgyegu’, in the dataset. Figure 1(a) shows the spatial distribution of Seoul's population at 9:00 A.M. on April 1, 2019 by the area unit.

Since the ‘jibgyegu’ area units are irregular in shape and size, we spatially aggregated the number of people recorded each hour in each area unit on a regular square grid of cell size 400 m. As a result, we obtained 4043 grid cells with geographic coordinates and hourly population density. We chose a grid cell size of 400 m to get a sufficient number of grid cells for statistical analysis but also to minimize the impact of noises from too-small size and the impact of natural boundaries such as mountains and rivers in Seoul. We also tested the robustness of our results with two different sizes of grid cells (250 and 500 m) and obtained qualitatively similar results. We considered weekdays in April 2019 for our analysis (20 days in total). 

To focus on relatively high population areas in cities, we excluded grid cells whose daily population densities are lower than the daily population densities averaged over all the grid cells. Figure 1(b) shows the spatial distribution of Seoul's population at 9:00 A.M. on April 1, 2019 on the grid cells after the exclusion of lowly-populated cells. The number of grid cells was reduced from 4043 to 2480. The excluded lowly-populated grid cells are mostly located in mountains, rivers, and public facilities such as sewage treatment plants.

\begin{figure}[htbp]
    \centering
    \includegraphics[width=9cm]{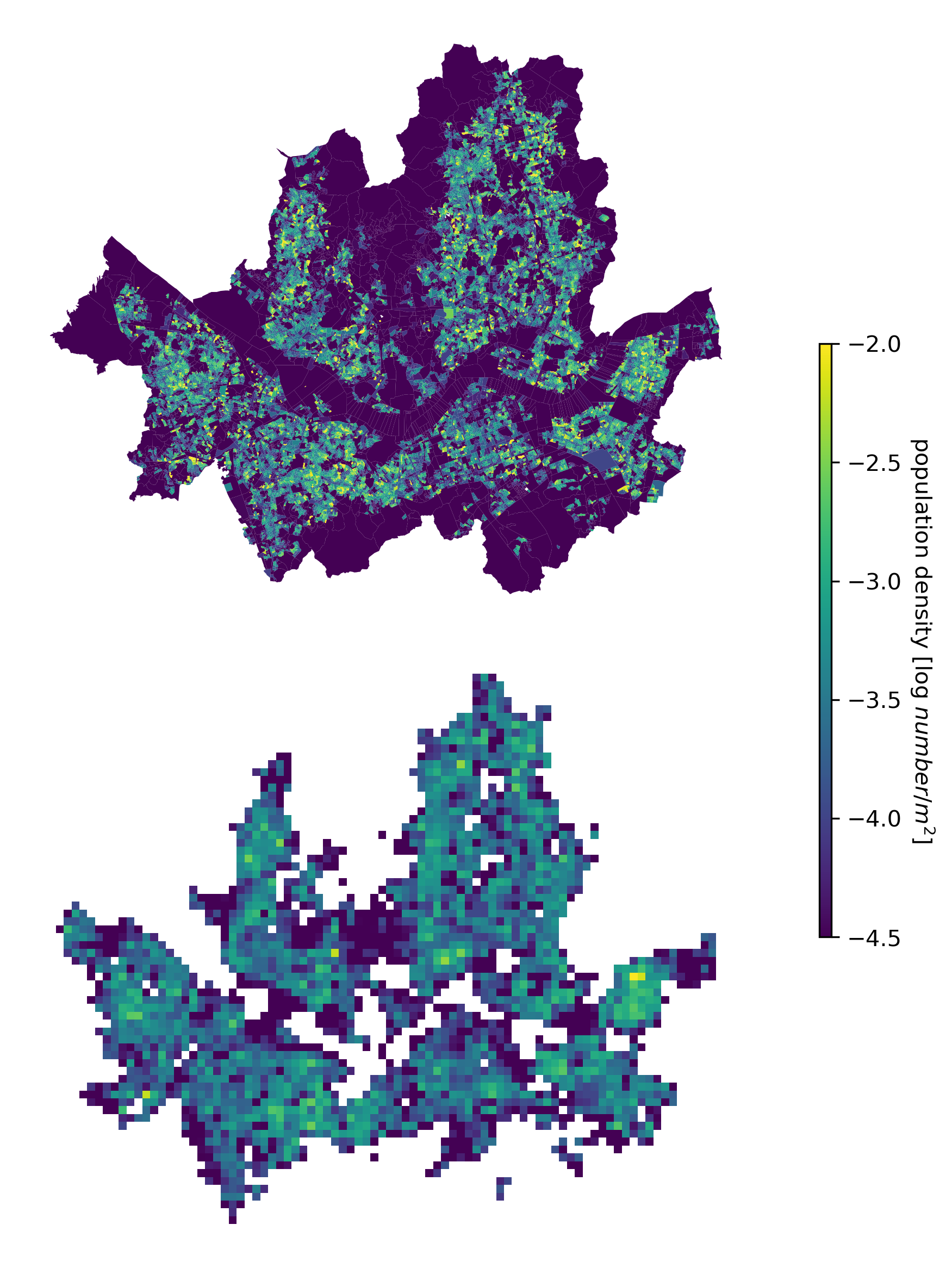}
    \caption{Spatial distribution of population density in Seoul at 9:00 A.M. on April 1, 2019. The color bar represents the logarithmic values of population density in spatial units. (a) Spatial distribution by the ‘jibgyegu’ units. (b) Spatial distribution by grid cells of 400 m size after the exclusion of lowly-populated cells.  }
    \label{fig:figure3}
\end{figure}

\subsection{\label{app:subsec} Population-based percolation}
To identify the macroscopic structure underlying the spatial distribution of urban population, we apply population-based percolation to the gridded population data. By occupying the grid cells one by one, from densely populated to sparsely populated, we can identify clusters of densely populated grid cells (the core parts of cities) and the connectivity between them.

The procedure of our population-based percolation is as follows. We first rank all the grid cells according to their population density from the highest to the lowest value. Next, we occupy the grid cells one by one according to their rank and observe how clusters evolve with the occupation probability $p$. A cluster, defined as a set of nearest neighboring occupied grid cells, represents a densely populated macroscopic area in a city. The occupation probability $p$ is defined as the number of occupied grid cells divided by the total number of grid cells.

\begin{figure*}
    \centering
    \includegraphics[width=\textwidth]{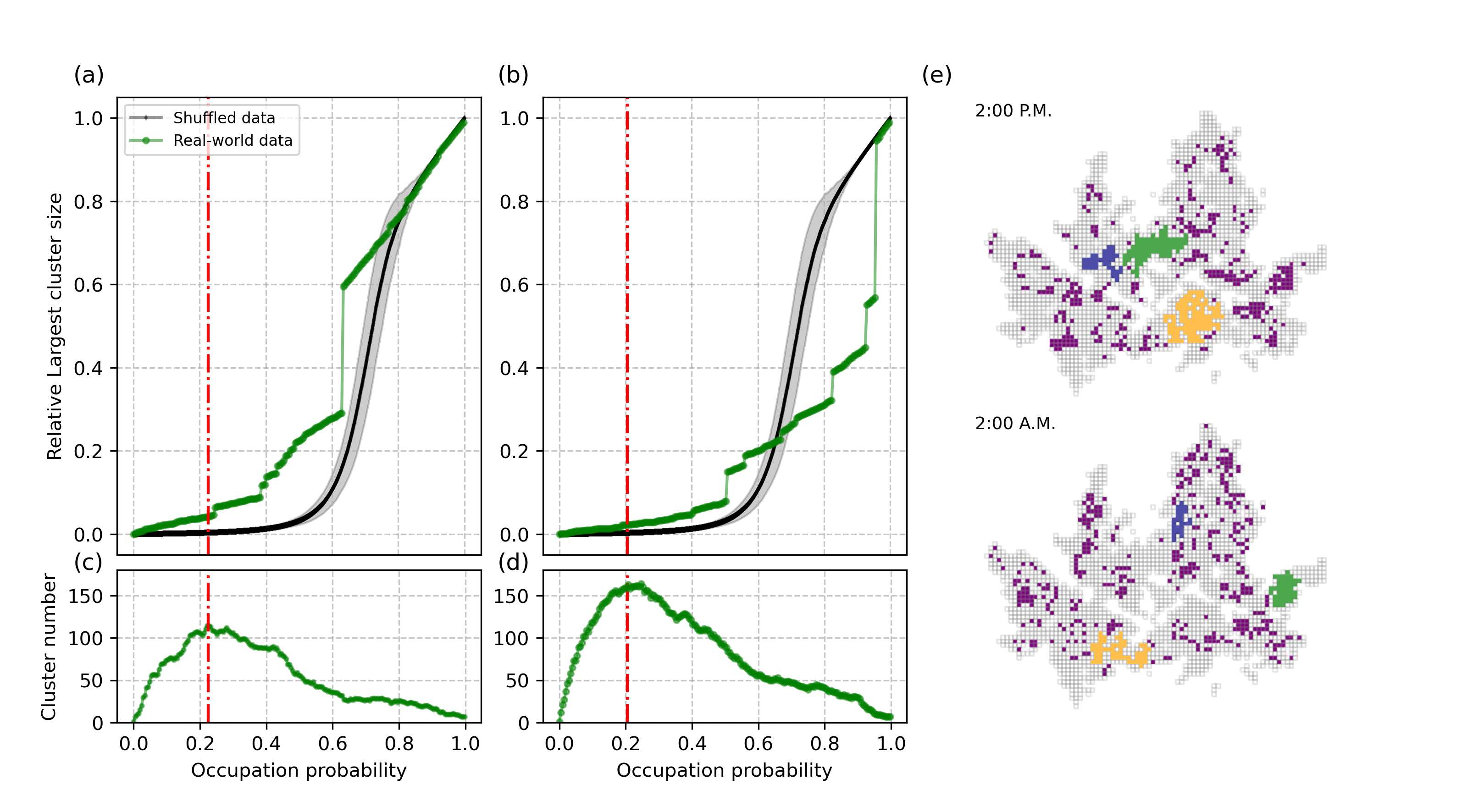}
    \caption{ The relative sizes of the largest cluster as a function of $p$ at (a) 2:00 P.M. and (b) 2:00 A.M. on Apr. 1st, 2019 in Seoul. Note that the relative size was obtained by dividing the cluster size by the total system size (i.e., 2480 grid cells).  Black lines show the relative size of the largest clusters in the case that occupation is based on the randomly shuffled population data. The number of clusters as a function of $p$  at (c) 2:00 P.M. and (d) 2:00 A.M. on Apr. 1st, 2019 in Seoul. Red dashdot lines is the number of clusters reaches its maximum, representing the critical occupation probability $p_c$. (e) Spatial distributions of clusters when $p=p_c$ at 2:00 P.M. and 2:00 A.M. on Apr. 1st, 2019 in Seoul.  We differently colored the top-ranked-size clusters : The yellow cluster represents the first largest cluster. The green cluster represents the second largest cluster. The blue cluster represents the third largest cluster. }
\end{figure*}

\section{Results}
\subsection{\label{app:subsec} Evolution of the largest cluster and the number of clusters}
Our percolation analysis aims at characterizing the macroscopic structure underlying spatial distribution of urban population by investigating the clusters of highly-populated areas in cities. First we investigate the size of the largest cluster as a function of the occupation probability $p$. Figure 2(a) and 2(b) show the size of the largest cluster at 2:00 P.M. and 2:00 A.M. on April 1, 2019 in Seoul as a function of $p$, respectively.

Black lines in figure 2(a) and 2(b) show the size of the largest clusters in the case that occupation is based on the random shuffling of the real-world population data. Such occupation is expected to be equivalent to the classical uncorrelated random site percolation with a continuous phase transition at $p \sim 0.59$. The evolution patterns of the size of the largest cluster in the real-world data and the shuffled data are quite different. The largest cluster in the real-world data starts to grow progressively from $p=0$ and exhibit intermittent jumps, whereas the largest cluster in the shuffled data is very small before a critical occupation probability and then suddenly grow, as predicted by the random percolation theory~\cite{stauffer2018introduction,christensen2005complexity}. These results indicate that in the real-world data, high-population grid cells are more clustered than they would be in the shuffled data and the largest cluster grows with merging other clusters in the real-world data, whereas it suddenly emerges in the shuffled data.

The evolution patterns of the size of the largest cluster between daytime (e.g., 2:00 P.M. in Fig. 2(a)) and nighttime (e.g., 2:00 A.M. in Fig. 2(b)) are also different. Compared to the size of the largest cluster at 2:00 P.M., the size of the largest cluster at 2:00 A.M. grows more slowly in the low occupation probability yet more frequently jumps. These results indicate that high-population grid cells during daytime are more clustered than during nighttime and cluster merging occurs more often during nighttime than during daytime.
%To examine the heterogeneity of spatial population distribution using another method,%

Next, we measure the number of clusters~\cite{ambuhl2023understanding, shreevastava2019emergent} in the real-world data as a function of $p$. Figure 2(c) and Figure 2(d) show the number of clusters as a function of $p$ at 2:00 P.M. and 2:00 A.M. on Apr. 1st, 2019 in Seoul, respectively. When $p$ is low, the number of clusters is increasing, indicating that high density grid cells create multiple clusters. After the number of cluster reaches its maximum (the red dashdot lines in Fig. 2), clusters tend to merge to the largest cluster and the largest cluster starts to grow rapidly. We define the critical occupation probability $p_c$ as the value of occupation probability when the number of clusters reaches its maximum. If maximum value of the number of clusters occurs at multiple occupation probabilities, we define $p_c$ as the value of occupation probability when the number of clusters first reaches its maximum. Figure 2(e) illustrates the spatial distribution of clusters when $p=p_c$ at 2:00 P.M. and 2:00 A.M. on April 1st, 2019.

\subsection{\label{app:subsec} Radius of gyration and correlation length }
We measure the radius of gyration of occupied cells and correlation length of clusters in the real-world data at each occupation probability $p$ to quantitatively understand the spatial range of of highly populated cells and the cluster size, respectively.

The radius of gyration $R_g(p)$ is defined as $R_{g}(p)$ = $\frac{1}{m}\sum_{i=1}^{m}\left|r_{i}-r_{cm}\right|^{2}$ where $r_{i}$ denotes the position of the occupied grid cell $i$, $r_{cm}$ is the center of mass of the occupied grid cells and defined as $r_{cm}$ = $\frac{1}{m}\sum_{i=1}^{m} r_{i}$, and $m$ is the total number of the occupied grid cell for a given $p$. $R_g(p)$ represents the typical length scale of the system of occupied grid cells at given $p$, indicating how much the occupied grid cells are dispersed from their center of mass at $p$.

The correlation length \(\xi\) is defined as the average distance of grid cells belonging to the same cluster, $\xi^2(p)$ =$\sum_{s}^{'}2R_{s}^{2} s^{2} n_{s}(p) / \sum_{s}^{'}s^{2} n_{s}(p)$, where $n_{s}(p)$ denotes the number of clusters of size $s$ at occupation probability $p$, $R_{s}$ is the radius of gyration of a given cluster with size $s$, and the prime on the sums indicates the exclusion of the largest cluster in each measurement~\cite{ali2013percolation}.

\begin{figure}[htbp]
    \centering
    \includegraphics[width=9cm]{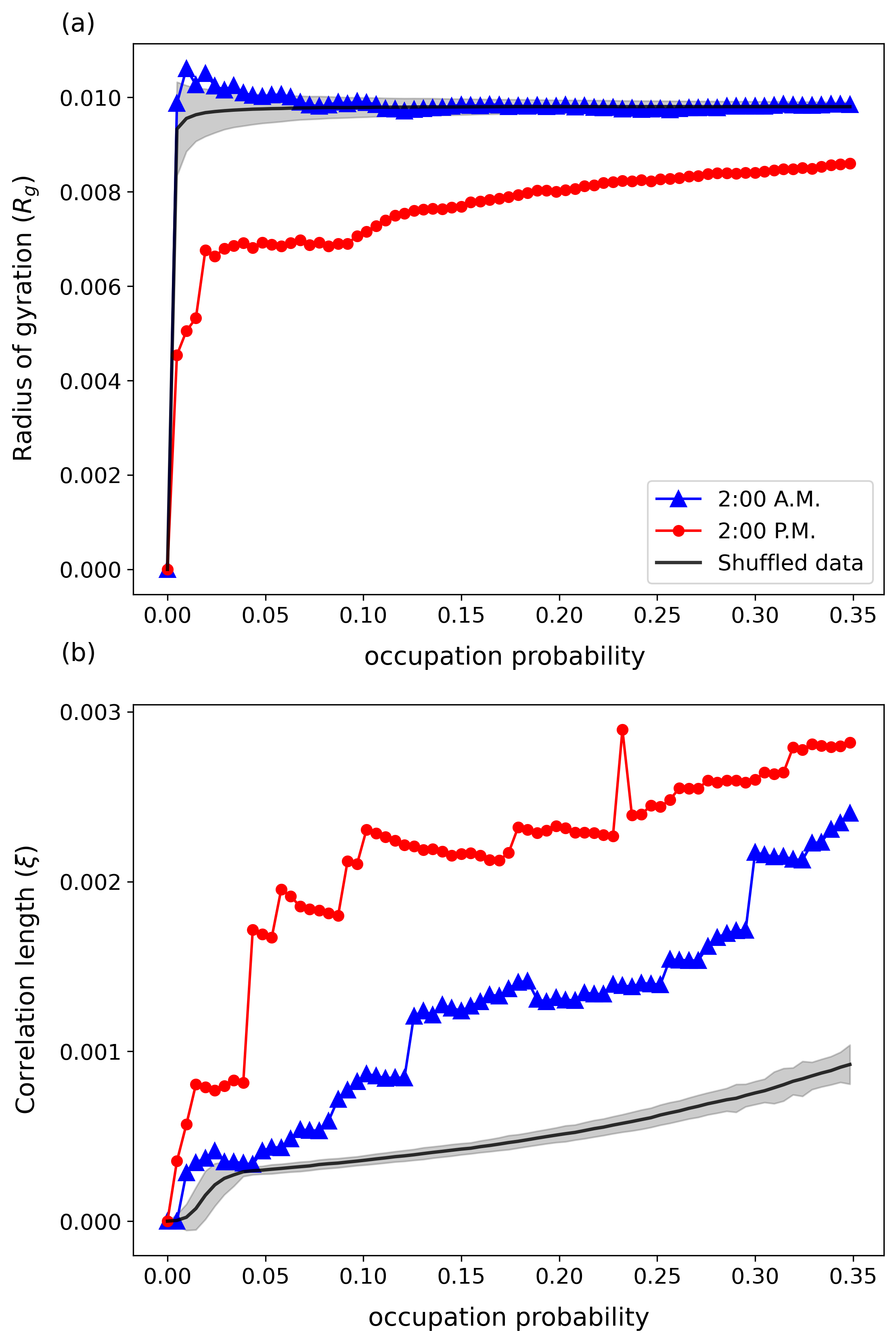}
    \caption{Radius of gyration $R_g(p)$ of occupied grid cells and correlation length $\xi(p)$ of clusters as a function of $p$ on April 1st, 2019 in Seoul. Both $R_g(p)$ and $\xi(p)$ are expressed in units relative to the system size, which is defined as the total number of grid cells in the system. (a) Radius of gyration $R_{g}(p)$. (b) Correlation length $\xi(p)$.
    }
    \label{fig:figure3}
\end{figure}

Figure 3(a) and 3(b) show the $R_{g}(p)$ and $\xi(p)$ at 2:00 P.M. and 2:00 A.M. on April 1st, 2019 in Seoul as a function of $p$, respectively. We represented the interval of low occupation probability (i.e., $0.0 \le p \le 0.35$) to focus on the behavior of high-density grid cells.

Figure 3(a) shows the $R_{g}(p)$ at 2:00 P.M. is smaller than the $R_{g}(p)$ at 2:00 A.M. This indicates that high-density grid cells at 2:00 P.M. are more spatially packed than those at 2:00 A.M. 

At 2:00 P.M., when the occupation probability ranges from $p$ = 0 to $p$ = 0.1, $R_{g}(p)$ remains stable and relatively small after the initial rapid growth, indicating that the top 10$\%$ of population density cells are located within a limited area from the center of mass. From $p$ = 0.1, $R_{g}(p)$ shows a monotonic increase pattern, implying that, as the population density of the grid cells decreases, they are located further away from their center of mass. 

On the other hand, at 2:00 A.M., when the occupation probability ranges from $p=0$ to $p=0.1$, $R_g(p)$ decreases from the larger radius than at 2:00 P.M., indicating that the top 10$\%$ of population density cells are more dispersed than at 2:00 P.M. But, as the population density of the grid cells decreases, they are located closer to their center of mass. From $p=0.1$ to $p=0.35$, $R_g(p)$ remains stable, implying that cell with population density from top $10$$\%$ to $35$$\%$ are similarly dispersed from the center of mass.

Figure 3(b) shows the $\xi(p)$ increases at 2:00 P.M and 2:00 A.M., as $p$ increases, indicating the typical size of clusters of high-density grid cells grows with $p$. The $\xi(p)$ observed at 2:00 P.M. is larger than at 2:00 A.M., indicating the presence of the larger clusters at 2:00 P.M. compared to 2:00 A.M. 

Black lines in figure 3(a) and 3(b) show the $R_g(p)$ and $\xi(p)$ in the case where occupation is based on the random shuffling of the real-world population data, respectively. We observe that $R_g(p)$ at 2:00 P.M. is lower than $R_g(p)$ in the shuffled data. On the other hand, $R_g(p)$ at 2:00 A.M. is slightly higher than $R_g(p)$ in the shuffled data when $p$ is low, but as $p$ increases, the two become comparable. These results indicate that high-population grid cells during daytime are more spatially clustered than those in the shuffled data, and high-population grid cells during nighttime are similarly dispersed compared to the shuffled data. We observe that $\xi(p)$ at 2:00 PM and at 2:00 AM are larger than the $\xi(p)$ in the shuffled data, indicating the typical size of clusters of high-density grid cells at 2:00 PM and at 2:00 AM are larger than in the shuffled data.

We considered these quantities for other times of the day, and found similar results: high-density grid cells during the daytime are more packed and form larger clusters than high-density grid cells during the nighttime.

\subsection{\label{app:subsec} Critical exponent of cluster size distribution $\tau$ and fractal dimension $D_{f}$}
To investigate whether there are fundamental and qualitative differences between spatial distribution of urban population between daytime and nighttime, we measure the critical exponent $\tau$ of cluster size distribution~\cite{ebrahimabadi2023geometry, zeng2019switch}. At the critical occupation probability $p_c$, where the total number of clusters is at the maximum, it is suggested that the cluster size distribution of finite clusters follows a power law :

\begin{equation}
{n(s)}  \sim
 s^{-\tau}
 \label{eq:wideeq}
\end{equation}

Here, $s$ is the cluster size, $n(s)$ is the total number of s-sized clusters, and $\tau$ is the corresponding critical exponent. 

Figure 4(a) shows the probability density function of population cluster size distribution at the critical point, following a power law, for 2:00 P.M. and 2:00 A.M. on April 1st, 2019. We find that the critical exponent $\tau$  = 1.89 at 2:00 P.M. is smaller than that the critical exponent $\tau$ = 2.28 at 2:00 A.M. 

As shown in Fig. 2(e), the sizes of large clusters are bigger and there are less small-size clusters at 2:00 P.M. than at 2:00 A.M., leading to a more skewed tail in the size distribution. In other words, cluster size distribution at criticality during daytime are more heterogeneous than during nighttime.

Different values of the critical exponent $\tau$ during daytime and nighttime suggest that spatial distribution of urban population during daytime and nighttime can belong to different universal classes. To check the validity of this result, we calculated the critical exponent $\tau$ at 2:00 P.M. and 2:00 A.M. for 20 weekdays in April 2019 and observed that, albeit with fluctuations, the values of $\tau$ at 2:00 P.M. tend to be distributed around 2.2 yet the values of $\tau$ at 2:00 A.M. tend to be distributed around 1.9 as shown in Fig. 4(b).

\begin{figure}[htbp]
    \centering
    \includegraphics[width=9cm]{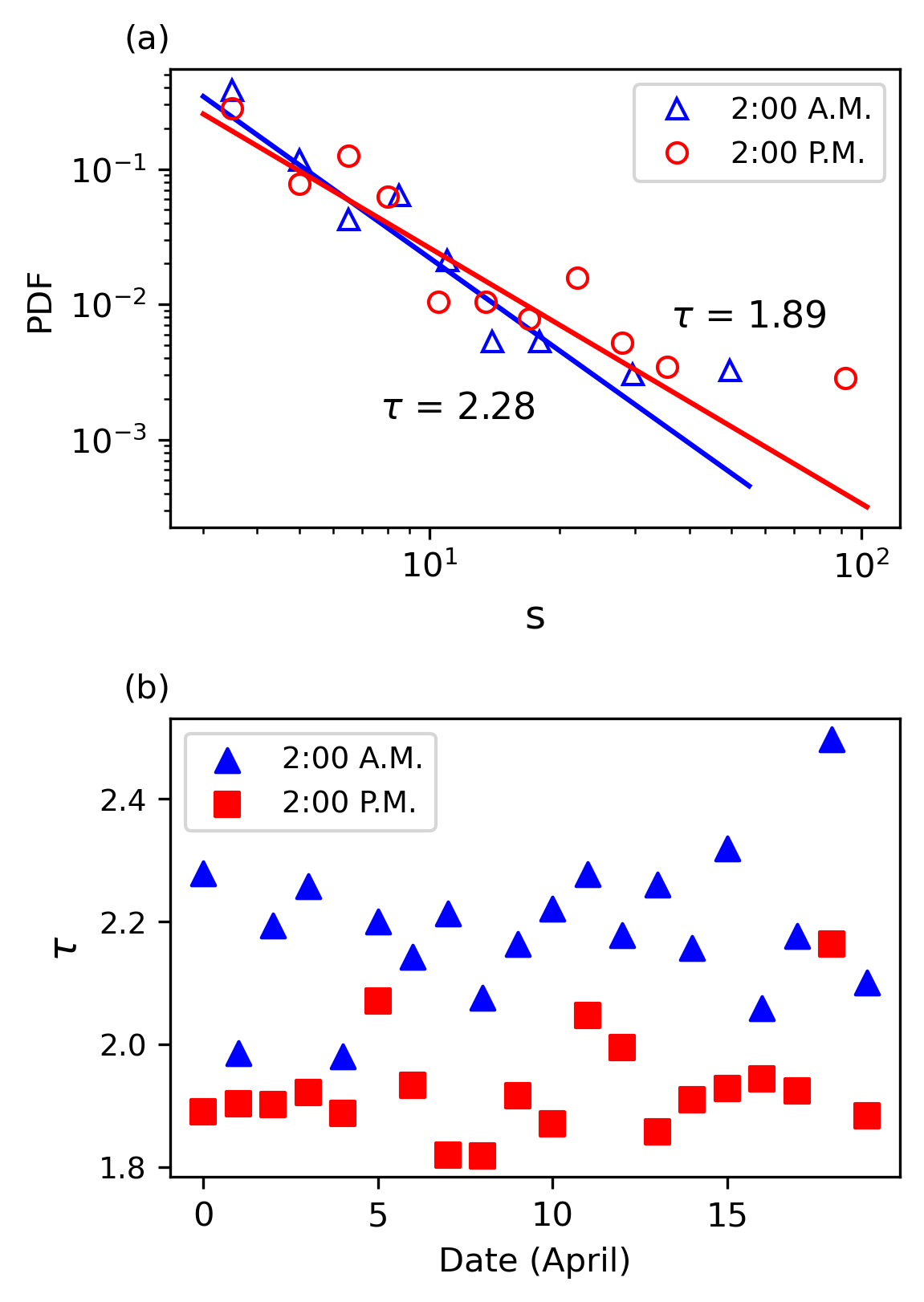}
    \caption{ (a) Cluster size distributions at percolation criticality and their power-law scaling exponents at 2:00 P.M.(red circle) and 2:00 A.M.(blue triangle) on Apr 1st, 2019 in Seoul. Lines indicate the scaling exponent $\tau$ = 1.89 at 2:00 P.M.(red) and $\tau$ = 2.28 at 2:00 A.M.(blue), respectively. (b) $\tau$ during workdays (Monday through Friday) on Apr, 2019 in Seoul (20 days). }
    \label{fig:figure4}
\end{figure}

A previous study on a city growth model using correlated percolation has shown that the exponent of the power-law distribution of urban settlement areas is influenced by long-range correlations between grid cells~\cite{makse1998modeling}. Applying this to our results, we can infer that as long-range correlations become stronger, there is a higher probability of larger clusters occurring, resulting in a decrease in the $\tau$ value. Consequently, we can hypothesize that there might be a stronger long-range correlation in the spatial distribution of daytime population than in that of nighttime population.

To check the validity of the measured critical exponent $\tau$, we employ maximum likelihood fitting methods with goodness-of-fit tests based on the Kolmogorov-Smirnov (KS) statistic and likelihood ratios~\cite{clauset2009power}. When performing a goodness-of-fit test on the real data, we obtain p-values of 0.35 for 2:00 A.M. and 0.10 for 2:00 P.M. These results suggest that it is reasonable to conclude that the distribution follows a power-law behavior in both cases.

The power-law form of cluster size distribution indicates the fractal structure of clusters. Scale-invariant and statistically self-similar clusters are observed in various phenomena from geographical objects such as lakes and islands~\cite{imre2016fractals} to urban heat islands~\cite{shreevastava2019emergent}. To quantitatively characterize geometrical differences in the clusters during the daytime and nighttime, we measure the aggregated area - perimeter fractal dimension of the collection of cluster at each occupation probability $p$ using the following equation:

\begin{equation}
{ \sum_{i=1}^{n_{p}} P_{i} \sim  \sum_{i=1}^{n_{p}} A_{i}^{\frac{D_{f}}{2}}, }
 \label{eq:wideeq}
\end{equation}

where $n_{p}$ denotes the total number of clusters at a given occupation probability $p$, and $P_{i}$ and $A_{i}$ are the perimeter and the area of cluster $i$. We define the perimeter of a cluster as the whole boundary between the cluster and the adjacent nonoccupied grid cells. The fractal dimension $D_f$ typically ranges between 1 and 2. $D_{f}$ approach 2 when clusters have the highest degree of irregularity (meaning that the perimeter fills the plane in the cluster) and approach 1 when clusters have a smooth shape such as a circle~\cite{shreevastava2019emergent}. It is also known that, for statistically self-similar surfaces, $D_f$ has fractional values between 1 and 2 and these values are independent of the occupation probability used for clustering~\cite{shreevastava2019emergent,isichenko1991statistical}.

\begin{figure}[htbp]
    \centering
    \includegraphics[width=9cm]{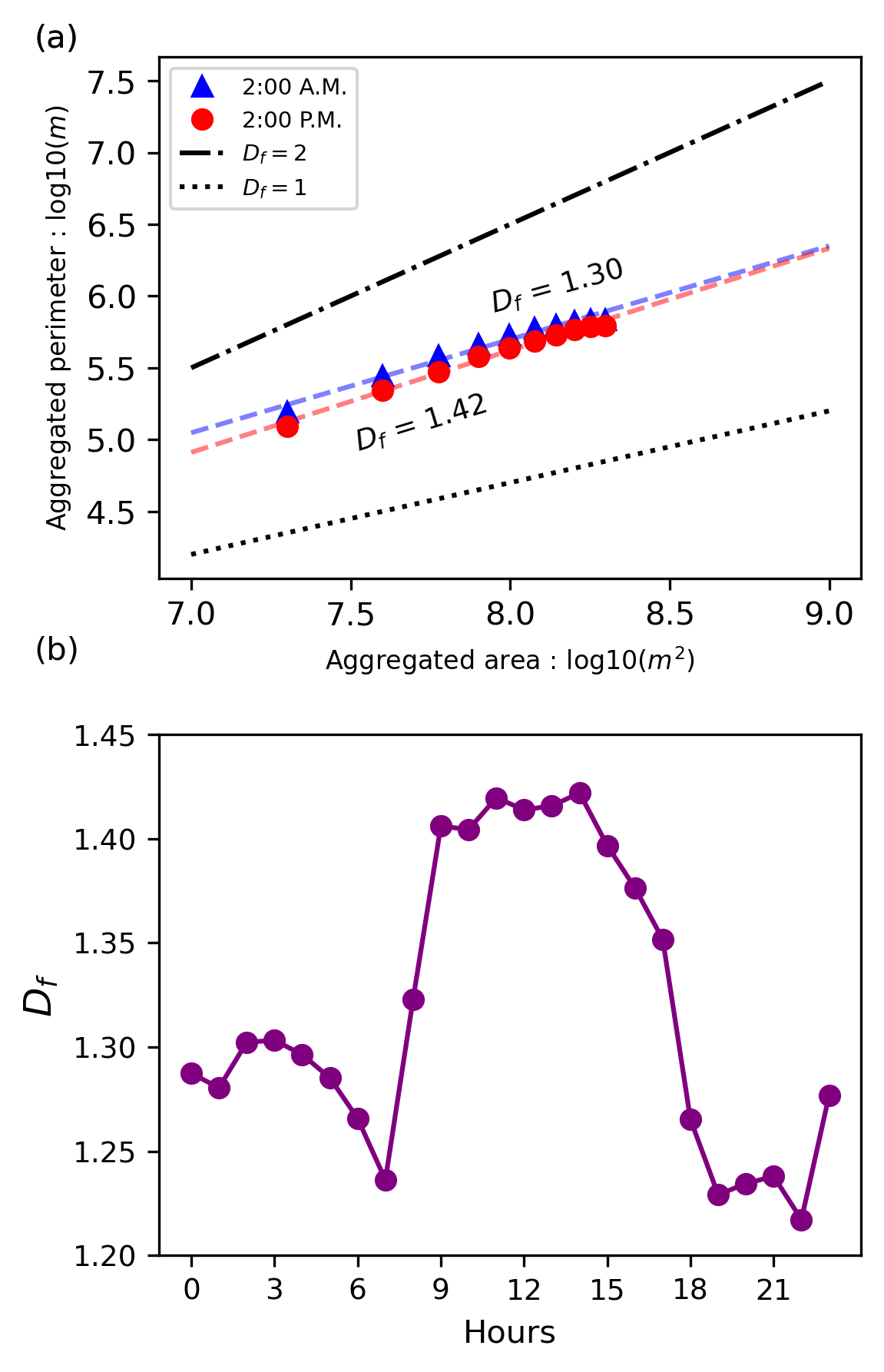}
    \caption{(a) Aggregated perimeters versus aggregated areas of clusters for occupation probabilities ranging from 0.05 to 0.5, with intervals of 0.05, at 2:00 P.M.(red) and 2:00 A.M.(blue) on April 1st, 2019, demonstrating the same ratio of log(Area) and log(Perimeter). Dashed lines show ratio $D_{f}$ = 1.42 and $D_{f}$ = 1.30 in the log - log plot. $D_{f}$ values of the perimeter of a circle ($D_{f}$ = 1) and a space-filling plane ($D_{f}$ = 2) are plotted to show the physical bounds for $D_f$. (b) Variation of $D_{f}$ April 1st, 2019.}
\end{figure}

Figure 5(a) shows log - log plots of the aggregated area-perimeter of clusters for multiple occupation probability of $p$ (0.05, 0.1, $\ldots$, 0.5). The data at 2:00 P.M. and 2:00 A.M. exhibit consistent ratios between log(Area) and log(Perimeter length), respectively. We observe that $D_f$=$1.42$ at 2:00 P.M. but $D_f$=$1.30$ at 2:00 A.M. Figure 5(b) shows the hourly variations in the fractal dimension ($D_{f}$) on April 1, 2019. Overall, we can observe that $D_f$ during the daytime is higher than that during the nighttime.

The reason for these differences of $D_{f}$ is understood by the morphology of large clusters (e.g., the large clusters colored yellow, green, and blue in Fig. 2(e)). As shown in Fig. 2(e), these large clusters are more irregular in shape and there are more internal voids in the largest cluster(colored yellow) at 2:00 P.M. than at 2:00 A.M.

\subsection{\label{app:subsec} Percolation analysis of spatial distribution of population in Helsinki}

\begin{figure*}
    \centering
    \includegraphics[width=\textwidth]{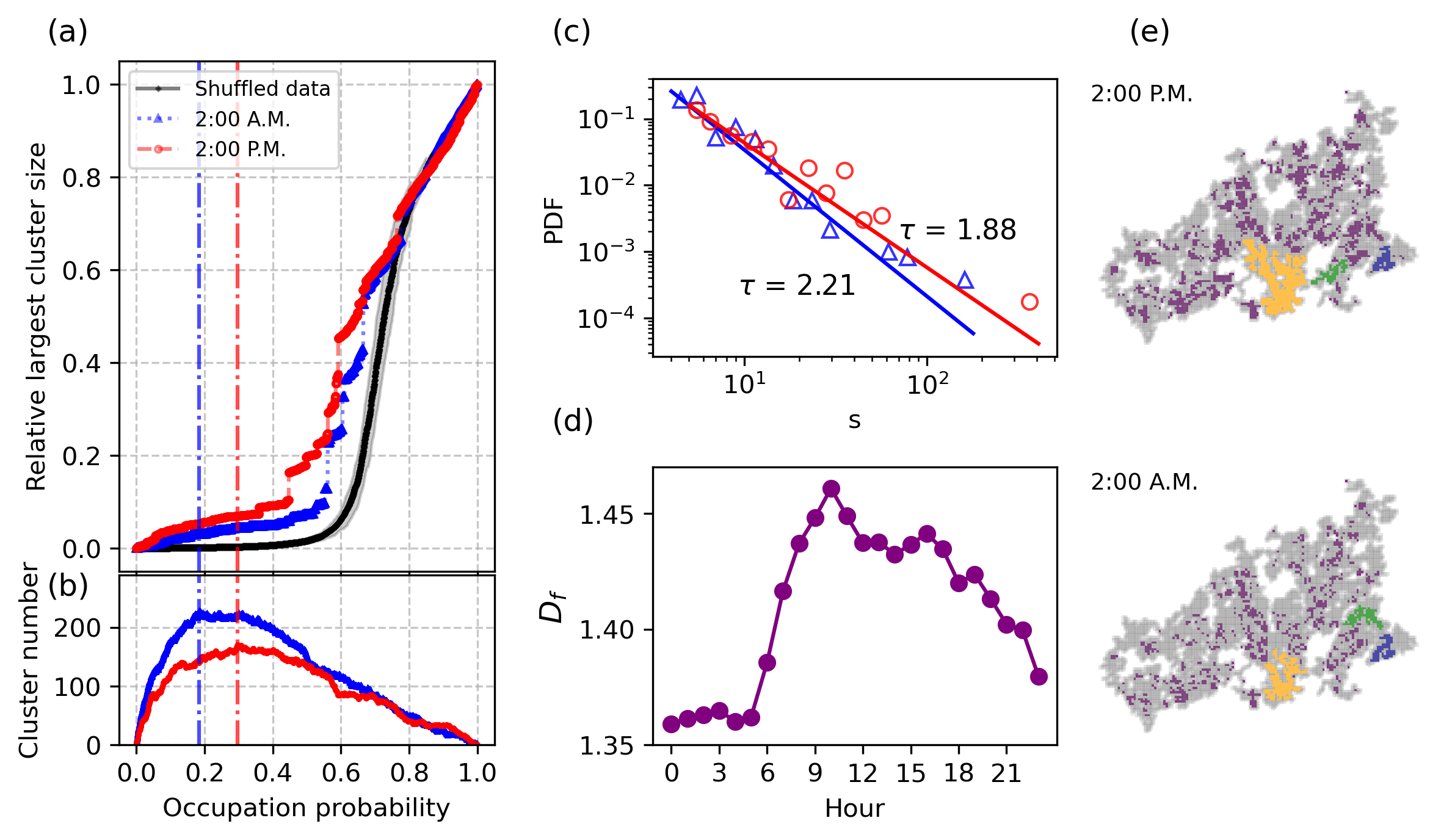}
    \caption{ { Results of applying population-based percolation to Helsinki population data (a) Evolution of the largest cluster relative size as a function of $p$ at 2:00 P.M.(red) and 2:00 A.M.(blue). Black lines show the relative size of the largest clusters in the case that occupation is based on the random shuffling of Helsinki population data. (b) Evolution of the number of clusters as a function of $p$ at 2:00 P.M.(red) and 2:00 A.M.(blue). Dashdot lines is number of cluster reaches its maximum, defined as the critical point $p_{c}$. (c) Cluster size distributions at percolation criticality and their power-law scaling exponents at 2:00 P.M.(red circle) and 2:00 A.M.(blue triangle) on workdays. Lines indicate scaling exponent $\tau$ = 1.88 at 2:00 P.M.(red) and $\tau$ = 2.21 at 2:00 A.M.(blue), respectively. (d) Variation of $D_{f}$ during the day. (e)  Snapshots of clusters at the critical point for 2:00 P.M. and 2:00 A.M. In the snapshot, we differently colored the top-ranked-size clusters: The yellow cluster represents the first largest cluster. The green cluster represents the second largest cluster. The blue cluster represents the third largest cluster.}}
\end{figure*}

To check the validity of our percolation framework for other cities, we consider Helsinki, the capital of Finland. We used the dynamic population distribution dataset for Helsinki Metropolitan Area(DHMA) obtained from Ref.~\cite{bergroth202224}, which is represented by 250 m x 250 m statistical grid cells. This dataset provides hourly relative population distribution data for workdays (From Monday to Thursday). Since the urban structure, grid cell size, and geographical features of Helsinki differ from those of Seoul, applying our percolation framework to DHMA dataset helps to support the validity of our percolation framework.

First we investigate the size of the largest cluster at 2:00 P.M. and 2:00 A.M. on workdays in Helsinki as a function of occupation probability $p$, respectively as shown in Fig 6(a). Similar to the results obtained from Seoul's data, the largest clusters in Helsinki start to grow progressively at low occupation probability $p$ and exhibit intermittent jumps. The evolution pattern of the largest cluster size is also different between daytime (2:00 P.M.) and night time (2:00 A.M.). Figure 6(b) show the number of clusters as a function of increasing $p$. The number of clusters increase up to the critical occupation probability $p_c$ and decrease thereafter as observed in Seoul.

In Figure 6(c), we measure the critical exponent $\tau$ of cluster size distribution at percolation criticality. We observe that $\tau=1.88$ at 2:00 P.M. but $\tau=2.21$ at 2:00 A.M. The pattern that the critical exponent $\tau$ is smaller during daytime than during nighttime is consistent with observations from the Seoul data. 

We also measure the fractal dimension $D_{f}$ of the collection of clusters for workdays in Helsinki. We observe that $D_{f}=1.43$ at 2:00 P.M. but $D_{f}=1.36$ at 2:00 A.M. In Figure 6(d), the pattern that fractal dimension $D_{f}$ is higher during daytime than during nighttime is consistent with observations from the Seoul data. 

The obtained values of $D_f$ in Helsinki tend to be higher than in Seoul. This is likely due to the fact that Helsinki has more forests and lakes than Seoul, and it is adjacent to the sea, thus the boundaries of where people can live are more rough in Helsinki than in Seoul. The critical exponent $\tau$ and fractal dimension $D_f$ of Seoul on April 1, 2019 and Helsinki on workdays are summarized in Table I for comparison.

\begin{table}[h]
\begin{ruledtabular}
\begin{tabular}{lcc}
& \textrm{Critical exponent} $\tau$ & \shortstack{\textrm{Area-perimeter}\\\textrm{fractal dimension}} \\
\colrule
Seoul (2:00 P.M.) & 1.89 & 1.42 \\
Seoul (2:00 A.M.) & 2.28 & 1.30 \\
Helsinki (2:00 P.M.) & 1.88 & 1.43 \\
Helsinki (2:00 A.M.) & 2.21 & 1.36 \\
\end{tabular}
\end{ruledtabular}
\caption{\label{tab:table1}%
Critical exponent $\tau$ and fractal dimension $D_f$ for the considered cities and time.
}
\label{tab:critical_exponent}
\end{table}

\section{Discussion} 
In summary, we identified the macroscopic structure underlying the spatial distribution of near-real time population in Seoul and Helsinki with a percolation framework. We observe that the growth patterns of the largest clusters in the real population data and their randomized data are significantly different, and highly populated areas during daytime are denser and form larger clusters than during nighttime. We investigated the cluster size distributions at percolation criticality and observed that their power-law exponents during the daytime are lower than during the nighttime, indicating that spatial distributions of urban population during daytime and nighttime fall into different universality class. We measured the area-perimeter fractal dimension of the collection of clusters and observed that the fractal dimensions during daytime are higher than during nighttime, indicating that the perimeters of high-population clusters during daytime are more rough than during nighttime. Our findings suggest that even the same city can have qualitatively different spatial distributions of population over time, and propose a way to quantitatively compare the macrostructure of cities based on population distribution data.

 What makes such qualitative differences between daytime and nighttime population distributions in the same city? A possible answer to this question is commuting as people concentrate in a few small areas (e.g., urban centers) during daytime and disperse after work. However, more quantitative analysis based on commuting data is necessary to check the validity of this hypothesis and this will be one of immediate future works of our study. Mathematical models such as fractional Brownian motion models may help address this question as in the other real-world application of percolation~\cite{saberi2020evidence}. Active-particle models can also be a good candidate for future research on the emergence of the observed patterns in the spatial distributions of urban population. Finally, one can investigate population distributions in other cities with different geometries, geographies, and socio-economic conditions with our framework to check whether there is a universal pattern across cities or whether cities can be classified into different groups~\cite{ebrahimabadi2023geometry}. Comparing our results from mobile phone users to the results from the registered population collected by census will be also interesting as nighttime population distribution is often assumed to be similar to the registered population distribution. 

 We argue that our framework minimizes the influence of microscopic details of the system such as grid cell size since it focuses on the critical exponent and the fractal dimension. But our framework assumes that population density is spatially heterogeneous and high-density areas are clustered in cities. In addition, if the population of a city is too sparely distributed in space, population clusters are not well defined. Our percolation-based analysis of macrostructure of dynamic urban population can provide an informative way to quantitatively understand the location of populations and compare the spatial distribution of population across cities or over time. Our findings suggest that urban planners and policy makers need to consider the differences in demand between daytime and nighttime when allocating resources and infrastructure in cities.

\begin{acknowledgments}
This work was supported by the Basic Study and Interdisciplinary R$\&$D Foundation Fund of the University of Seoul (2022).
% We wish to acknowledge the support of the author community in using
% REV\TeX{}, offering suggestions and encouragement, testing new versions,
% \dots.
\end{acknowledgments}

\bibliography{CSNS2408}  %

\begin{thebibliography}{41}%
\makeatletter
\providecommand \@ifxundefined [1]{%
 \@ifx{#1\undefined}
}%
\providecommand \@ifnum [1]{%
 \ifnum #1\expandafter \@firstoftwo
 \else \expandafter \@secondoftwo
 \fi
}%
\providecommand \@ifx [1]{%
 \ifx #1\expandafter \@firstoftwo
 \else \expandafter \@secondoftwo
 \fi
}%
\providecommand \natexlab [1]{#1}%
\providecommand \enquote  [1]{``#1''}%
\providecommand \bibnamefont  [1]{#1}%
\providecommand \bibfnamefont [1]{#1}%
\providecommand \citenamefont [1]{#1}%
\providecommand \href@noop [0]{\@secondoftwo}%
\providecommand \href [0]{\begingroup \@sanitize@url \@href}%
\providecommand \@href[1]{\@@startlink{#1}\@@href}%
\providecommand \@@href[1]{\endgroup#1\@@endlink}%
\providecommand \@sanitize@url [0]{\catcode `\\12\catcode `\$12\catcode `\&12\catcode `\#12\catcode `\^12\catcode `\_12\catcode `\%12\relax}%
\providecommand \@@startlink[1]{}%
\providecommand \@@endlink[0]{}%
\providecommand \url  [0]{\begingroup\@sanitize@url \@url }%
\providecommand \@url [1]{\endgroup\@href {#1}{\urlprefix }}%
\providecommand \urlprefix  [0]{URL }%
\providecommand \Eprint [0]{\href }%
\providecommand \doibase [0]{https://doi.org/}%
\providecommand \selectlanguage [0]{\@gobble}%
\providecommand \bibinfo  [0]{\@secondoftwo}%
\providecommand \bibfield  [0]{\@secondoftwo}%
\providecommand \translation [1]{[#1]}%
\providecommand \BibitemOpen [0]{}%
\providecommand \bibitemStop [0]{}%
\providecommand \bibitemNoStop [0]{.\EOS\space}%
\providecommand \EOS [0]{\spacefactor3000\relax}%
\providecommand \BibitemShut  [1]{\csname bibitem#1\endcsname}%
\let\auto@bib@innerbib\@empty
%</preamble>
\bibitem [{\citenamefont {Volpati}\ and\ \citenamefont {Barthelemy}(2018)}]{volpati2018spatial}%
  \BibitemOpen
  \bibfield  {author} {\bibinfo {author} {\bibfnamefont {V.}~\bibnamefont {Volpati}}\ and\ \bibinfo {author} {\bibfnamefont {M.}~\bibnamefont {Barthelemy}},\ }\bibfield  {title} {\bibinfo {title} {The spatial organization of the population density in cities},\ }\href@noop {} {\bibfield  {journal} {\bibinfo  {journal} {arXiv preprint arXiv:1804.00855}\ } (\bibinfo {year} {2018})}\BibitemShut {NoStop}%
\bibitem [{\citenamefont {Schwarz}(2010)}]{schwarz2010urban}%
  \BibitemOpen
  \bibfield  {author} {\bibinfo {author} {\bibfnamefont {N.}~\bibnamefont {Schwarz}},\ }\bibfield  {title} {\bibinfo {title} {Urban form revisited—selecting indicators for characterising european cities},\ }\href@noop {} {\bibfield  {journal} {\bibinfo  {journal} {Landscape and urban planning}\ }\textbf {\bibinfo {volume} {96}},\ \bibinfo {pages} {29} (\bibinfo {year} {2010})}\BibitemShut {NoStop}%
\bibitem [{\citenamefont {Barthelemy}(2016)}]{barthelemy2016structure}%
  \BibitemOpen
  \bibfield  {author} {\bibinfo {author} {\bibfnamefont {M.}~\bibnamefont {Barthelemy}},\ }\href@noop {} {\emph {\bibinfo {title} {The Structure and Dynamics of Cities: Urban Data Analysis and Theoretical Modeling}}}\ (\bibinfo  {publisher} {Cambridge University Press},\ \bibinfo {address} {Cambridge, UK},\ \bibinfo {year} {2016})\BibitemShut {NoStop}%
\bibitem [{\citenamefont {Bettencourt}(2021)}]{bettencourt2021introduction}%
  \BibitemOpen
  \bibfield  {author} {\bibinfo {author} {\bibfnamefont {L.~M.~A.}\ \bibnamefont {Bettencourt}},\ }\href {https://doi.org/10.7551/mitpress/13761.001.0001} {\emph {\bibinfo {title} {Introduction to Urban Science: Evidence and Theory of Cities as Complex Systems}}}\ (\bibinfo  {publisher} {MIT Press},\ \bibinfo {address} {Cambridge, MA},\ \bibinfo {year} {2021})\BibitemShut {NoStop}%
\bibitem [{\citenamefont {Hong}\ \emph {et~al.}(2022)\citenamefont {Hong}, \citenamefont {Hui},\ and\ \citenamefont {Lin}}]{hong2022relationship}%
  \BibitemOpen
  \bibfield  {author} {\bibinfo {author} {\bibfnamefont {S.}~\bibnamefont {Hong}}, \bibinfo {author} {\bibfnamefont {E.~C.-m.}\ \bibnamefont {Hui}},\ and\ \bibinfo {author} {\bibfnamefont {Y.}~\bibnamefont {Lin}},\ }\bibfield  {title} {\bibinfo {title} {Relationship between urban spatial structure and carbon emissions: A literature review},\ }\href@noop {} {\bibfield  {journal} {\bibinfo  {journal} {Ecological Indicators}\ }\textbf {\bibinfo {volume} {144}},\ \bibinfo {pages} {109456} (\bibinfo {year} {2022})}\BibitemShut {NoStop}%
\bibitem [{\citenamefont {Xu}\ \emph {et~al.}(2023)\citenamefont {Xu}, \citenamefont {Olmos}, \citenamefont {Mateo}, \citenamefont {Hernando}, \citenamefont {Yang},\ and\ \citenamefont {Gonz{\'a}lez}}]{xu2023urban}%
  \BibitemOpen
  \bibfield  {author} {\bibinfo {author} {\bibfnamefont {Y.}~\bibnamefont {Xu}}, \bibinfo {author} {\bibfnamefont {L.~E.}\ \bibnamefont {Olmos}}, \bibinfo {author} {\bibfnamefont {D.}~\bibnamefont {Mateo}}, \bibinfo {author} {\bibfnamefont {A.}~\bibnamefont {Hernando}}, \bibinfo {author} {\bibfnamefont {X.}~\bibnamefont {Yang}},\ and\ \bibinfo {author} {\bibfnamefont {M.~C.}\ \bibnamefont {Gonz{\'a}lez}},\ }\bibfield  {title} {\bibinfo {title} {Urban dynamics through the lens of human mobility},\ }\href@noop {} {\bibfield  {journal} {\bibinfo  {journal} {Nature computational science}\ }\textbf {\bibinfo {volume} {3}},\ \bibinfo {pages} {611} (\bibinfo {year} {2023})}\BibitemShut {NoStop}%
\bibitem [{\citenamefont {Sang}\ \emph {et~al.}(2024)\citenamefont {Sang}, \citenamefont {Jiang}, \citenamefont {Huang}, \citenamefont {Zhu},\ and\ \citenamefont {Wang}}]{sang2024spatial}%
  \BibitemOpen
  \bibfield  {author} {\bibinfo {author} {\bibfnamefont {M.}~\bibnamefont {Sang}}, \bibinfo {author} {\bibfnamefont {J.}~\bibnamefont {Jiang}}, \bibinfo {author} {\bibfnamefont {X.}~\bibnamefont {Huang}}, \bibinfo {author} {\bibfnamefont {F.}~\bibnamefont {Zhu}},\ and\ \bibinfo {author} {\bibfnamefont {Q.}~\bibnamefont {Wang}},\ }\bibfield  {title} {\bibinfo {title} {Spatial and temporal changes in population distribution and population projection at county level in china},\ }\href@noop {} {\bibfield  {journal} {\bibinfo  {journal} {Humanities and Social Sciences Communications}\ }\textbf {\bibinfo {volume} {11}},\ \bibinfo {pages} {1} (\bibinfo {year} {2024})}\BibitemShut {NoStop}%
\bibitem [{\citenamefont {Crosato}\ \emph {et~al.}(2021)\citenamefont {Crosato}, \citenamefont {Prokopenko},\ and\ \citenamefont {Harr{\'e}}}]{crosato2021polycentric}%
  \BibitemOpen
  \bibfield  {author} {\bibinfo {author} {\bibfnamefont {E.}~\bibnamefont {Crosato}}, \bibinfo {author} {\bibfnamefont {M.}~\bibnamefont {Prokopenko}},\ and\ \bibinfo {author} {\bibfnamefont {M.~S.}\ \bibnamefont {Harr{\'e}}},\ }\bibfield  {title} {\bibinfo {title} {The polycentric dynamics of melbourne and sydney: Suburb attractiveness divides a city at the home ownership level},\ }\href@noop {} {\bibfield  {journal} {\bibinfo  {journal} {Proceedings of the Royal Society A}\ }\textbf {\bibinfo {volume} {477}},\ \bibinfo {pages} {20200514} (\bibinfo {year} {2021})}\BibitemShut {NoStop}%
\bibitem [{\citenamefont {Dong}\ \emph {et~al.}(2020)\citenamefont {Dong}, \citenamefont {Huang}, \citenamefont {Zhang},\ and\ \citenamefont {Liu}}]{dong2020understanding}%
  \BibitemOpen
  \bibfield  {author} {\bibinfo {author} {\bibfnamefont {L.}~\bibnamefont {Dong}}, \bibinfo {author} {\bibfnamefont {Z.}~\bibnamefont {Huang}}, \bibinfo {author} {\bibfnamefont {J.}~\bibnamefont {Zhang}},\ and\ \bibinfo {author} {\bibfnamefont {Y.}~\bibnamefont {Liu}},\ }\bibfield  {title} {\bibinfo {title} {Understanding the mesoscopic scaling patterns within cities},\ }\href@noop {} {\bibfield  {journal} {\bibinfo  {journal} {Scientific reports}\ }\textbf {\bibinfo {volume} {10}},\ \bibinfo {pages} {21201} (\bibinfo {year} {2020})}\BibitemShut {NoStop}%
\bibitem [{\citenamefont {Louail}\ \emph {et~al.}(2015)\citenamefont {Louail}, \citenamefont {Lenormand}, \citenamefont {Picornell}, \citenamefont {Garcia~Cantu}, \citenamefont {Herranz}, \citenamefont {Frias-Martinez}, \citenamefont {Ramasco},\ and\ \citenamefont {Barthelemy}}]{louail2015uncovering}%
  \BibitemOpen
  \bibfield  {author} {\bibinfo {author} {\bibfnamefont {T.}~\bibnamefont {Louail}}, \bibinfo {author} {\bibfnamefont {M.}~\bibnamefont {Lenormand}}, \bibinfo {author} {\bibfnamefont {M.}~\bibnamefont {Picornell}}, \bibinfo {author} {\bibfnamefont {O.}~\bibnamefont {Garcia~Cantu}}, \bibinfo {author} {\bibfnamefont {R.}~\bibnamefont {Herranz}}, \bibinfo {author} {\bibfnamefont {E.}~\bibnamefont {Frias-Martinez}}, \bibinfo {author} {\bibfnamefont {J.~J.}\ \bibnamefont {Ramasco}},\ and\ \bibinfo {author} {\bibfnamefont {M.}~\bibnamefont {Barthelemy}},\ }\bibfield  {title} {\bibinfo {title} {Uncovering the spatial structure of mobility networks},\ }\href@noop {} {\bibfield  {journal} {\bibinfo  {journal} {Nature Communications}\ }\textbf {\bibinfo {volume} {6}},\ \bibinfo {pages} {6007} (\bibinfo {year} {2015})}\BibitemShut {NoStop}%
\bibitem [{\citenamefont {Lee}\ \emph {et~al.}(2021)\citenamefont {Lee}, \citenamefont {Goh}, \citenamefont {Lee},\ and\ \citenamefont {Choi}}]{lee2021spatiotemporal}%
  \BibitemOpen
  \bibfield  {author} {\bibinfo {author} {\bibfnamefont {J.-H.}\ \bibnamefont {Lee}}, \bibinfo {author} {\bibfnamefont {S.}~\bibnamefont {Goh}}, \bibinfo {author} {\bibfnamefont {K.}~\bibnamefont {Lee}},\ and\ \bibinfo {author} {\bibfnamefont {M.~Y.}\ \bibnamefont {Choi}},\ }\bibfield  {title} {\bibinfo {title} {Spatiotemporal distributions of population in seoul: joint influence of ridership and accessibility of the subway system},\ }\href@noop {} {\bibfield  {journal} {\bibinfo  {journal} {EPJ data science}\ }\textbf {\bibinfo {volume} {10}},\ \bibinfo {pages} {41} (\bibinfo {year} {2021})}\BibitemShut {NoStop}%
\bibitem [{\citenamefont {Wang}\ \emph {et~al.}(2009)\citenamefont {Wang}, \citenamefont {Gonz{\'a}lez}, \citenamefont {Hidalgo},\ and\ \citenamefont {Barab{\'a}si}}]{wang2009understanding}%
  \BibitemOpen
  \bibfield  {author} {\bibinfo {author} {\bibfnamefont {P.}~\bibnamefont {Wang}}, \bibinfo {author} {\bibfnamefont {M.~C.}\ \bibnamefont {Gonz{\'a}lez}}, \bibinfo {author} {\bibfnamefont {C.~A.}\ \bibnamefont {Hidalgo}},\ and\ \bibinfo {author} {\bibfnamefont {A.-L.}\ \bibnamefont {Barab{\'a}si}},\ }\bibfield  {title} {\bibinfo {title} {Understanding the spreading patterns of mobile phone viruses},\ }\href@noop {} {\bibfield  {journal} {\bibinfo  {journal} {Science}\ }\textbf {\bibinfo {volume} {324}},\ \bibinfo {pages} {1071} (\bibinfo {year} {2009})}\BibitemShut {NoStop}%
\bibitem [{\citenamefont {Lenormand}\ \emph {et~al.}(2014)\citenamefont {Lenormand}, \citenamefont {Picornell}, \citenamefont {Cant{\'u}-Ros}, \citenamefont {Tugores}, \citenamefont {Louail}, \citenamefont {Herranz}, \citenamefont {Barthelemy}, \citenamefont {Frias-Martinez},\ and\ \citenamefont {Ramasco}}]{lenormand2014cross}%
  \BibitemOpen
  \bibfield  {author} {\bibinfo {author} {\bibfnamefont {M.}~\bibnamefont {Lenormand}}, \bibinfo {author} {\bibfnamefont {M.}~\bibnamefont {Picornell}}, \bibinfo {author} {\bibfnamefont {O.~G.}\ \bibnamefont {Cant{\'u}-Ros}}, \bibinfo {author} {\bibfnamefont {A.}~\bibnamefont {Tugores}}, \bibinfo {author} {\bibfnamefont {T.}~\bibnamefont {Louail}}, \bibinfo {author} {\bibfnamefont {R.}~\bibnamefont {Herranz}}, \bibinfo {author} {\bibfnamefont {M.}~\bibnamefont {Barthelemy}}, \bibinfo {author} {\bibfnamefont {E.}~\bibnamefont {Frias-Martinez}},\ and\ \bibinfo {author} {\bibfnamefont {J.~J.}\ \bibnamefont {Ramasco}},\ }\bibfield  {title} {\bibinfo {title} {Cross-checking different sources of mobility information},\ }\href@noop {} {\bibfield  {journal} {\bibinfo  {journal} {PloS one}\ }\textbf {\bibinfo {volume} {9}},\ \bibinfo {pages} {e105184} (\bibinfo {year} {2014})}\BibitemShut {NoStop}%
\bibitem [{\citenamefont {Wu}\ \emph {et~al.}(2021)\citenamefont {Wu}, \citenamefont {Frias-Martinez},\ and\ \citenamefont {Frias-Martinez}}]{wu2021spatial}%
  \BibitemOpen
  \bibfield  {author} {\bibinfo {author} {\bibfnamefont {J.}~\bibnamefont {Wu}}, \bibinfo {author} {\bibfnamefont {E.}~\bibnamefont {Frias-Martinez}},\ and\ \bibinfo {author} {\bibfnamefont {V.}~\bibnamefont {Frias-Martinez}},\ }\bibfield  {title} {\bibinfo {title} {Spatial sensitivity analysis for urban hotspots using cell phone traces},\ }\href@noop {} {\bibfield  {journal} {\bibinfo  {journal} {Environment and Planning B: Urban Analytics and City Science}\ }\textbf {\bibinfo {volume} {48}},\ \bibinfo {pages} {2623} (\bibinfo {year} {2021})}\BibitemShut {NoStop}%
\bibitem [{\citenamefont {Louail}\ \emph {et~al.}(2014)\citenamefont {Louail}, \citenamefont {Lenormand}, \citenamefont {Cantu~Ros}, \citenamefont {Picornell}, \citenamefont {Herranz}, \citenamefont {Frias-Martinez}, \citenamefont {Ramasco},\ and\ \citenamefont {Barthelemy}}]{louail2014mobile}%
  \BibitemOpen
  \bibfield  {author} {\bibinfo {author} {\bibfnamefont {T.}~\bibnamefont {Louail}}, \bibinfo {author} {\bibfnamefont {M.}~\bibnamefont {Lenormand}}, \bibinfo {author} {\bibfnamefont {O.~G.}\ \bibnamefont {Cantu~Ros}}, \bibinfo {author} {\bibfnamefont {M.}~\bibnamefont {Picornell}}, \bibinfo {author} {\bibfnamefont {R.}~\bibnamefont {Herranz}}, \bibinfo {author} {\bibfnamefont {E.}~\bibnamefont {Frias-Martinez}}, \bibinfo {author} {\bibfnamefont {J.~J.}\ \bibnamefont {Ramasco}},\ and\ \bibinfo {author} {\bibfnamefont {M.}~\bibnamefont {Barthelemy}},\ }\bibfield  {title} {\bibinfo {title} {From mobile phone data to the spatial structure of cities},\ }\href@noop {} {\bibfield  {journal} {\bibinfo  {journal} {Scientific reports}\ }\textbf {\bibinfo {volume} {4}},\ \bibinfo {pages} {5276} (\bibinfo {year} {2014})}\BibitemShut {NoStop}%
\bibitem [{\citenamefont {Roth}\ \emph {et~al.}(2011)\citenamefont {Roth}, \citenamefont {Kang}, \citenamefont {Batty},\ and\ \citenamefont {Barth{\'e}lemy}}]{roth2011structure}%
  \BibitemOpen
  \bibfield  {author} {\bibinfo {author} {\bibfnamefont {C.}~\bibnamefont {Roth}}, \bibinfo {author} {\bibfnamefont {S.~M.}\ \bibnamefont {Kang}}, \bibinfo {author} {\bibfnamefont {M.}~\bibnamefont {Batty}},\ and\ \bibinfo {author} {\bibfnamefont {M.}~\bibnamefont {Barth{\'e}lemy}},\ }\bibfield  {title} {\bibinfo {title} {Structure of urban movements: polycentric activity and entangled hierarchical flows},\ }\href@noop {} {\bibfield  {journal} {\bibinfo  {journal} {PloS one}\ }\textbf {\bibinfo {volume} {6}},\ \bibinfo {pages} {e15923} (\bibinfo {year} {2011})}\BibitemShut {NoStop}%
\bibitem [{\citenamefont {Bassolas}\ \emph {et~al.}(2019)\citenamefont {Bassolas}, \citenamefont {Barbosa-Filho}, \citenamefont {Dickinson}, \citenamefont {Dotiwalla}, \citenamefont {Eastham}, \citenamefont {Gallotti}, \citenamefont {Ghoshal}, \citenamefont {Gipson}, \citenamefont {Hazarie}, \citenamefont {Kautz} \emph {et~al.}}]{bassolas2019hierarchical}%
  \BibitemOpen
  \bibfield  {author} {\bibinfo {author} {\bibfnamefont {A.}~\bibnamefont {Bassolas}}, \bibinfo {author} {\bibfnamefont {H.}~\bibnamefont {Barbosa-Filho}}, \bibinfo {author} {\bibfnamefont {B.}~\bibnamefont {Dickinson}}, \bibinfo {author} {\bibfnamefont {X.}~\bibnamefont {Dotiwalla}}, \bibinfo {author} {\bibfnamefont {P.}~\bibnamefont {Eastham}}, \bibinfo {author} {\bibfnamefont {R.}~\bibnamefont {Gallotti}}, \bibinfo {author} {\bibfnamefont {G.}~\bibnamefont {Ghoshal}}, \bibinfo {author} {\bibfnamefont {B.}~\bibnamefont {Gipson}}, \bibinfo {author} {\bibfnamefont {S.~A.}\ \bibnamefont {Hazarie}}, \bibinfo {author} {\bibfnamefont {H.}~\bibnamefont {Kautz}}, \emph {et~al.},\ }\bibfield  {title} {\bibinfo {title} {Hierarchical organization of urban mobility and its connection with city livability},\ }\href@noop {} {\bibfield  {journal} {\bibinfo  {journal} {Nature communications}\ }\textbf {\bibinfo {volume} {10}},\ \bibinfo {pages} {4817} (\bibinfo {year} {2019})}\BibitemShut {NoStop}%
\bibitem [{\citenamefont {Saberi}(2015)}]{saberi2015recent}%
  \BibitemOpen
  \bibfield  {author} {\bibinfo {author} {\bibfnamefont {A.~A.}\ \bibnamefont {Saberi}},\ }\bibfield  {title} {\bibinfo {title} {Recent advances in percolation theory and its applications},\ }\href@noop {} {\bibfield  {journal} {\bibinfo  {journal} {Physics Reports}\ }\textbf {\bibinfo {volume} {578}},\ \bibinfo {pages} {1} (\bibinfo {year} {2015})}\BibitemShut {NoStop}%
\bibitem [{\citenamefont {Stauffer}\ and\ \citenamefont {Aharony}(2018)}]{stauffer2018introduction}%
  \BibitemOpen
  \bibfield  {author} {\bibinfo {author} {\bibfnamefont {D.}~\bibnamefont {Stauffer}}\ and\ \bibinfo {author} {\bibfnamefont {A.}~\bibnamefont {Aharony}},\ }\href@noop {} {\emph {\bibinfo {title} {Introduction to percolation theory}}}\ (\bibinfo  {publisher} {Taylor \& Francis},\ \bibinfo {year} {2018})\BibitemShut {NoStop}%
\bibitem [{\citenamefont {Ebrahimabadi}\ \emph {et~al.}(2023)\citenamefont {Ebrahimabadi}, \citenamefont {Hosseiny}, \citenamefont {Fan},\ and\ \citenamefont {Saberi}}]{ebrahimabadi2023geometry}%
  \BibitemOpen
  \bibfield  {author} {\bibinfo {author} {\bibfnamefont {S.}~\bibnamefont {Ebrahimabadi}}, \bibinfo {author} {\bibfnamefont {A.}~\bibnamefont {Hosseiny}}, \bibinfo {author} {\bibfnamefont {J.}~\bibnamefont {Fan}},\ and\ \bibinfo {author} {\bibfnamefont {A.~A.}\ \bibnamefont {Saberi}},\ }\bibfield  {title} {\bibinfo {title} {Geometry of commutes in the universality of percolating traffic flows},\ }\href@noop {} {\bibfield  {journal} {\bibinfo  {journal} {Physical Review E}\ }\textbf {\bibinfo {volume} {108}},\ \bibinfo {pages} {054311} (\bibinfo {year} {2023})}\BibitemShut {NoStop}%
\bibitem [{\citenamefont {Zeng}\ \emph {et~al.}(2019)\citenamefont {Zeng}, \citenamefont {Li}, \citenamefont {Guo}, \citenamefont {Gao}, \citenamefont {Gao}, \citenamefont {Stanley},\ and\ \citenamefont {Havlin}}]{zeng2019switch}%
  \BibitemOpen
  \bibfield  {author} {\bibinfo {author} {\bibfnamefont {G.}~\bibnamefont {Zeng}}, \bibinfo {author} {\bibfnamefont {D.}~\bibnamefont {Li}}, \bibinfo {author} {\bibfnamefont {S.}~\bibnamefont {Guo}}, \bibinfo {author} {\bibfnamefont {L.}~\bibnamefont {Gao}}, \bibinfo {author} {\bibfnamefont {Z.}~\bibnamefont {Gao}}, \bibinfo {author} {\bibfnamefont {H.~E.}\ \bibnamefont {Stanley}},\ and\ \bibinfo {author} {\bibfnamefont {S.}~\bibnamefont {Havlin}},\ }\bibfield  {title} {\bibinfo {title} {Switch between critical percolation modes in city traffic dynamics},\ }\href@noop {} {\bibfield  {journal} {\bibinfo  {journal} {Proceedings of the National Academy of Sciences}\ }\textbf {\bibinfo {volume} {116}},\ \bibinfo {pages} {23} (\bibinfo {year} {2019})}\BibitemShut {NoStop}%
\bibitem [{\citenamefont {Kwon}\ \emph {et~al.}(2023)\citenamefont {Kwon}, \citenamefont {Jung},\ and\ \citenamefont {Eom}}]{kwon2023global}%
  \BibitemOpen
  \bibfield  {author} {\bibinfo {author} {\bibfnamefont {Y.}~\bibnamefont {Kwon}}, \bibinfo {author} {\bibfnamefont {J.-H.}\ \bibnamefont {Jung}},\ and\ \bibinfo {author} {\bibfnamefont {Y.-H.}\ \bibnamefont {Eom}},\ }\bibfield  {title} {\bibinfo {title} {Global efficiency and network structure of urban traffic flows: A percolation-based empirical analysis},\ }\href@noop {} {\bibfield  {journal} {\bibinfo  {journal} {Chaos: An Interdisciplinary Journal of Nonlinear Science}\ }\textbf {\bibinfo {volume} {33}},\ \bibinfo {pages} {113104} (\bibinfo {year} {2023})}\BibitemShut {NoStop}%
\bibitem [{\citenamefont {Cao}\ \emph {et~al.}(2020)\citenamefont {Cao}, \citenamefont {Dong}, \citenamefont {Wu},\ and\ \citenamefont {Liu}}]{cao2020quantifying}%
  \BibitemOpen
  \bibfield  {author} {\bibinfo {author} {\bibfnamefont {W.}~\bibnamefont {Cao}}, \bibinfo {author} {\bibfnamefont {L.}~\bibnamefont {Dong}}, \bibinfo {author} {\bibfnamefont {L.}~\bibnamefont {Wu}},\ and\ \bibinfo {author} {\bibfnamefont {Y.}~\bibnamefont {Liu}},\ }\bibfield  {title} {\bibinfo {title} {Quantifying urban areas with multi-source data based on percolation theory},\ }\href@noop {} {\bibfield  {journal} {\bibinfo  {journal} {Remote Sensing of Environment}\ }\textbf {\bibinfo {volume} {241}},\ \bibinfo {pages} {111730} (\bibinfo {year} {2020})}\BibitemShut {NoStop}%
\bibitem [{\citenamefont {Amb{\"u}hl}\ \emph {et~al.}(2023)\citenamefont {Amb{\"u}hl}, \citenamefont {Menendez},\ and\ \citenamefont {Gonz{\'a}lez}}]{ambuhl2023understanding}%
  \BibitemOpen
  \bibfield  {author} {\bibinfo {author} {\bibfnamefont {L.}~\bibnamefont {Amb{\"u}hl}}, \bibinfo {author} {\bibfnamefont {M.}~\bibnamefont {Menendez}},\ and\ \bibinfo {author} {\bibfnamefont {M.~C.}\ \bibnamefont {Gonz{\'a}lez}},\ }\bibfield  {title} {\bibinfo {title} {Understanding congestion propagation by combining percolation theory with the macroscopic fundamental diagram},\ }\href {https://doi.org/10.1038/s42005-023-01144-w} {\bibfield  {journal} {\bibinfo  {journal} {Communications Physics}\ }\textbf {\bibinfo {volume} {6}},\ \bibinfo {pages} {1} (\bibinfo {year} {2023})}\BibitemShut {NoStop}%
\bibitem [{\citenamefont {Ali~Saberi}(2013)}]{ali2013percolation}%
  \BibitemOpen
  \bibfield  {author} {\bibinfo {author} {\bibfnamefont {A.}~\bibnamefont {Ali~Saberi}},\ }\bibfield  {title} {\bibinfo {title} {Percolation description of the global topography of earth and the moon},\ }\href@noop {} {\bibfield  {journal} {\bibinfo  {journal} {Physical review letters}\ }\textbf {\bibinfo {volume} {110}},\ \bibinfo {pages} {178501} (\bibinfo {year} {2013})}\BibitemShut {NoStop}%
\bibitem [{\citenamefont {Fan}\ \emph {et~al.}(2019)\citenamefont {Fan}, \citenamefont {Meng},\ and\ \citenamefont {Saberi}}]{fan2019percolation}%
  \BibitemOpen
  \bibfield  {author} {\bibinfo {author} {\bibfnamefont {J.}~\bibnamefont {Fan}}, \bibinfo {author} {\bibfnamefont {J.}~\bibnamefont {Meng}},\ and\ \bibinfo {author} {\bibfnamefont {A.~A.}\ \bibnamefont {Saberi}},\ }\bibfield  {title} {\bibinfo {title} {Percolation framework of the earth's topography},\ }\href@noop {} {\bibfield  {journal} {\bibinfo  {journal} {Physical Review E}\ }\textbf {\bibinfo {volume} {99}},\ \bibinfo {pages} {022304} (\bibinfo {year} {2019})}\BibitemShut {NoStop}%
\bibitem [{\citenamefont {Sun}\ \emph {et~al.}(2021)\citenamefont {Sun}, \citenamefont {Meng}, \citenamefont {Yao}, \citenamefont {Saberi}, \citenamefont {Chen}, \citenamefont {Fan},\ and\ \citenamefont {Kurths}}]{sun2021percolation}%
  \BibitemOpen
  \bibfield  {author} {\bibinfo {author} {\bibfnamefont {Y.}~\bibnamefont {Sun}}, \bibinfo {author} {\bibfnamefont {J.}~\bibnamefont {Meng}}, \bibinfo {author} {\bibfnamefont {Q.}~\bibnamefont {Yao}}, \bibinfo {author} {\bibfnamefont {A.~A.}\ \bibnamefont {Saberi}}, \bibinfo {author} {\bibfnamefont {X.}~\bibnamefont {Chen}}, \bibinfo {author} {\bibfnamefont {J.}~\bibnamefont {Fan}},\ and\ \bibinfo {author} {\bibfnamefont {J.}~\bibnamefont {Kurths}},\ }\bibfield  {title} {\bibinfo {title} {Percolation analysis of the atmospheric structure},\ }\href@noop {} {\bibfield  {journal} {\bibinfo  {journal} {Physical Review E}\ }\textbf {\bibinfo {volume} {104}},\ \bibinfo {pages} {064139} (\bibinfo {year} {2021})}\BibitemShut {NoStop}%
\bibitem [{\citenamefont {Rozenfeld}\ \emph {et~al.}(2011)\citenamefont {Rozenfeld}, \citenamefont {Rybski}, \citenamefont {Gabaix},\ and\ \citenamefont {Makse}}]{rozenfeld2011area}%
  \BibitemOpen
  \bibfield  {author} {\bibinfo {author} {\bibfnamefont {H.~D.}\ \bibnamefont {Rozenfeld}}, \bibinfo {author} {\bibfnamefont {D.}~\bibnamefont {Rybski}}, \bibinfo {author} {\bibfnamefont {X.}~\bibnamefont {Gabaix}},\ and\ \bibinfo {author} {\bibfnamefont {H.~A.}\ \bibnamefont {Makse}},\ }\bibfield  {title} {\bibinfo {title} {The area and population of cities: New insights from a different perspective on cities},\ }\href@noop {} {\bibfield  {journal} {\bibinfo  {journal} {American Economic Review}\ }\textbf {\bibinfo {volume} {101}},\ \bibinfo {pages} {2205} (\bibinfo {year} {2011})}\BibitemShut {NoStop}%
\bibitem [{\citenamefont {Shreevastava}\ \emph {et~al.}(2019)\citenamefont {Shreevastava}, \citenamefont {Rao},\ and\ \citenamefont {McGrath}}]{shreevastava2019emergent}%
  \BibitemOpen
  \bibfield  {author} {\bibinfo {author} {\bibfnamefont {A.}~\bibnamefont {Shreevastava}}, \bibinfo {author} {\bibfnamefont {P.~S.~C.}\ \bibnamefont {Rao}},\ and\ \bibinfo {author} {\bibfnamefont {G.~S.}\ \bibnamefont {McGrath}},\ }\bibfield  {title} {\bibinfo {title} {Emergent self-similarity and scaling properties of fractal intra-urban heat islets for diverse global cities},\ }\href@noop {} {\bibfield  {journal} {\bibinfo  {journal} {Physical Review E}\ }\textbf {\bibinfo {volume} {100}},\ \bibinfo {pages} {032142} (\bibinfo {year} {2019})}\BibitemShut {NoStop}%
\bibitem [{\citenamefont {Deng}\ \emph {et~al.}(2019)\citenamefont {Deng}, \citenamefont {Liu}, \citenamefont {Liu},\ and\ \citenamefont {Luo}}]{deng2019detecting}%
  \BibitemOpen
  \bibfield  {author} {\bibinfo {author} {\bibfnamefont {Y.}~\bibnamefont {Deng}}, \bibinfo {author} {\bibfnamefont {J.}~\bibnamefont {Liu}}, \bibinfo {author} {\bibfnamefont {Y.}~\bibnamefont {Liu}},\ and\ \bibinfo {author} {\bibfnamefont {A.}~\bibnamefont {Luo}},\ }\bibfield  {title} {\bibinfo {title} {Detecting urban polycentric structure from poi data},\ }\href@noop {} {\bibfield  {journal} {\bibinfo  {journal} {ISPRS International Journal of Geo-Information}\ }\textbf {\bibinfo {volume} {8}},\ \bibinfo {pages} {283} (\bibinfo {year} {2019})}\BibitemShut {NoStop}%
\bibitem [{\citenamefont {Sadewo}\ \emph {et~al.}(2021)\citenamefont {Sadewo}, \citenamefont {Syabri}, \citenamefont {Antipova}, \citenamefont {Pradono},\ and\ \citenamefont {Hudalah}}]{sadewo2021using}%
  \BibitemOpen
  \bibfield  {author} {\bibinfo {author} {\bibfnamefont {E.}~\bibnamefont {Sadewo}}, \bibinfo {author} {\bibfnamefont {I.}~\bibnamefont {Syabri}}, \bibinfo {author} {\bibfnamefont {A.}~\bibnamefont {Antipova}}, \bibinfo {author} {\bibnamefont {Pradono}},\ and\ \bibinfo {author} {\bibfnamefont {D.}~\bibnamefont {Hudalah}},\ }\bibfield  {title} {\bibinfo {title} {Using morphological and functional polycentricity analyses to study the indonesian urban spatial structure: the case of medan, jakarta, and denpasar},\ }\href@noop {} {\bibfield  {journal} {\bibinfo  {journal} {Asian geographer}\ }\textbf {\bibinfo {volume} {38}},\ \bibinfo {pages} {47} (\bibinfo {year} {2021})}\BibitemShut {NoStop}%
\bibitem [{\citenamefont {Li}\ and\ \citenamefont {Liu}(2018)}]{li2018did}%
  \BibitemOpen
  \bibfield  {author} {\bibinfo {author} {\bibfnamefont {Y.}~\bibnamefont {Li}}\ and\ \bibinfo {author} {\bibfnamefont {X.}~\bibnamefont {Liu}},\ }\bibfield  {title} {\bibinfo {title} {How did urban polycentricity and dispersion affect economic productivity? a case study of 306 chinese cities},\ }\href@noop {} {\bibfield  {journal} {\bibinfo  {journal} {Landscape and Urban Planning}\ }\textbf {\bibinfo {volume} {173}},\ \bibinfo {pages} {51} (\bibinfo {year} {2018})}\BibitemShut {NoStop}%
\bibitem [{\citenamefont {Krehl}(2018)}]{krehl2018urban}%
  \BibitemOpen
  \bibfield  {author} {\bibinfo {author} {\bibfnamefont {A.}~\bibnamefont {Krehl}},\ }\bibfield  {title} {\bibinfo {title} {Urban subcentres in german city regions: Identification, understanding, comparison},\ }\href@noop {} {\bibfield  {journal} {\bibinfo  {journal} {Papers in Regional Science}\ }\textbf {\bibinfo {volume} {97}},\ \bibinfo {pages} {S79} (\bibinfo {year} {2018})}\BibitemShut {NoStop}%
\bibitem [{SLP()}]{SLPD}%
  \BibitemOpen
  \href@noop {} {\bibinfo {title} {{ Seoul Living Population Data}}},\ \bibinfo {howpublished} {\url{https://data.seoul.go.kr/dataVisual/seoul/seoulLivingPopulation.do}},\ \bibinfo {note} {2019}\BibitemShut {NoStop}%
\bibitem [{\citenamefont {Christensen}\ and\ \citenamefont {Moloney}(2005)}]{christensen2005complexity}%
  \BibitemOpen
  \bibfield  {author} {\bibinfo {author} {\bibfnamefont {K.}~\bibnamefont {Christensen}}\ and\ \bibinfo {author} {\bibfnamefont {N.}~\bibnamefont {Moloney}},\ }\href {https://books.google.co.kr/books?id=BM42DwAAQBAJ} {\emph {\bibinfo {title} {Complexity and Criticality}}},\ Imperial College Press Advanced Physics Texts\ (\bibinfo  {publisher} {World Scientific Publishing Company},\ \bibinfo {year} {2005})\BibitemShut {NoStop}%
\bibitem [{\citenamefont {Makse}\ \emph {et~al.}(1998)\citenamefont {Makse}, \citenamefont {Andrade}, \citenamefont {Batty}, \citenamefont {Havlin}, \citenamefont {Stanley} \emph {et~al.}}]{makse1998modeling}%
  \BibitemOpen
  \bibfield  {author} {\bibinfo {author} {\bibfnamefont {H.~A.}\ \bibnamefont {Makse}}, \bibinfo {author} {\bibfnamefont {J.~S.}\ \bibnamefont {Andrade}}, \bibinfo {author} {\bibfnamefont {M.}~\bibnamefont {Batty}}, \bibinfo {author} {\bibfnamefont {S.}~\bibnamefont {Havlin}}, \bibinfo {author} {\bibfnamefont {H.~E.}\ \bibnamefont {Stanley}}, \emph {et~al.},\ }\bibfield  {title} {\bibinfo {title} {Modeling urban growth patterns with correlated percolation},\ }\href@noop {} {\bibfield  {journal} {\bibinfo  {journal} {Physical Review E}\ }\textbf {\bibinfo {volume} {58}},\ \bibinfo {pages} {7054} (\bibinfo {year} {1998})}\BibitemShut {NoStop}%
\bibitem [{\citenamefont {Clauset}\ \emph {et~al.}(2009)\citenamefont {Clauset}, \citenamefont {Shalizi},\ and\ \citenamefont {Newman}}]{clauset2009power}%
  \BibitemOpen
  \bibfield  {author} {\bibinfo {author} {\bibfnamefont {A.}~\bibnamefont {Clauset}}, \bibinfo {author} {\bibfnamefont {C.~R.}\ \bibnamefont {Shalizi}},\ and\ \bibinfo {author} {\bibfnamefont {M.~E.}\ \bibnamefont {Newman}},\ }\bibfield  {title} {\bibinfo {title} {Power-law distributions in empirical data},\ }\href@noop {} {\bibfield  {journal} {\bibinfo  {journal} {SIAM review}\ }\textbf {\bibinfo {volume} {51}},\ \bibinfo {pages} {661} (\bibinfo {year} {2009})}\BibitemShut {NoStop}%
\bibitem [{\citenamefont {Imre}\ and\ \citenamefont {Novotn{\`y}}(2016)}]{imre2016fractals}%
  \BibitemOpen
  \bibfield  {author} {\bibinfo {author} {\bibfnamefont {A.~R.}\ \bibnamefont {Imre}}\ and\ \bibinfo {author} {\bibfnamefont {J.}~\bibnamefont {Novotn{\`y}}},\ }\bibfield  {title} {\bibinfo {title} {Fractals and the korcak-law: a history and a correction},\ }\href@noop {} {\bibfield  {journal} {\bibinfo  {journal} {The European Physical Journal H}\ }\textbf {\bibinfo {volume} {41}},\ \bibinfo {pages} {69} (\bibinfo {year} {2016})}\BibitemShut {NoStop}%
\bibitem [{\citenamefont {Isichenko}\ and\ \citenamefont {Kalda}(1991)}]{isichenko1991statistical}%
  \BibitemOpen
  \bibfield  {author} {\bibinfo {author} {\bibfnamefont {M.~B.}\ \bibnamefont {Isichenko}}\ and\ \bibinfo {author} {\bibfnamefont {J.}~\bibnamefont {Kalda}},\ }\bibfield  {title} {\bibinfo {title} {Statistical topography. i. fractal dimension of coastlines and number-area rule for islands},\ }\href@noop {} {\bibfield  {journal} {\bibinfo  {journal} {Journal of Nonlinear Science}\ }\textbf {\bibinfo {volume} {1}},\ \bibinfo {pages} {255} (\bibinfo {year} {1991})}\BibitemShut {NoStop}%
\bibitem [{\citenamefont {Bergroth}\ \emph {et~al.}(2022)\citenamefont {Bergroth}, \citenamefont {J{\"a}rv}, \citenamefont {Tenkanen}, \citenamefont {Manninen},\ and\ \citenamefont {Toivonen}}]{bergroth202224}%
  \BibitemOpen
  \bibfield  {author} {\bibinfo {author} {\bibfnamefont {C.}~\bibnamefont {Bergroth}}, \bibinfo {author} {\bibfnamefont {O.}~\bibnamefont {J{\"a}rv}}, \bibinfo {author} {\bibfnamefont {H.}~\bibnamefont {Tenkanen}}, \bibinfo {author} {\bibfnamefont {M.}~\bibnamefont {Manninen}},\ and\ \bibinfo {author} {\bibfnamefont {T.}~\bibnamefont {Toivonen}},\ }\bibfield  {title} {\bibinfo {title} {A 24-hour population distribution dataset based on mobile phone data from helsinki metropolitan area, finland},\ }\href@noop {} {\bibfield  {journal} {\bibinfo  {journal} {Scientific Data}\ }\textbf {\bibinfo {volume} {9}},\ \bibinfo {pages} {39} (\bibinfo {year} {2022})},\ \bibinfo {note} {\url{https://www.nature.com/articles/s41597-021-01113-4}}\BibitemShut {NoStop}%
\bibitem [{\citenamefont {Saberi}(2020)}]{saberi2020evidence}%
  \BibitemOpen
  \bibfield  {author} {\bibinfo {author} {\bibfnamefont {A.~A.}\ \bibnamefont {Saberi}},\ }\bibfield  {title} {\bibinfo {title} {Evidence for an ancient sea level on mars},\ }\href@noop {} {\bibfield  {journal} {\bibinfo  {journal} {The Astrophysical Journal Letters}\ }\textbf {\bibinfo {volume} {896}},\ \bibinfo {pages} {L25} (\bibinfo {year} {2020})}\BibitemShut {NoStop}%
\end{thebibliography}%
\end{document}